\documentstyle[12pt]{article}

\catcode`\@=11
\long\def\@makefntext#1{
\protect\noindent \hbox to 3.2pt {\hskip-.9pt
$^{{\ninerm\@thefnmark}}$\hfil}#1\hfill}                

\def\@makefnmark{\hbox to 0pt{$^{\@thefnmark}$\hss}}  

\def\ps@myheadings{\let\@mkboth\@gobbletwo
\def\@oddhead{\hbox{}
\rightmark\hfil\ninerm\thepage}
\def\@oddfoot{}\def\@evenhead{\ninerm\thepage\hfil
\leftmark\hbox{}}\def\@evenfoot{}
\def\sectionmark##1{}\def\subsectionmark##1{}}

\newcounter{sectionc}\newcounter{subsectionc}\newcounter{subsubsectionc}
\renewcommand{\section}[1] {\vspace*{0.6cm}\addtocounter{sectionc}{1}
\setcounter{subsectionc}{0}\setcounter{subsubsectionc}{0}\noindent
        {\normalsize\bf\thesectionc. #1}\par\vspace*{0.4cm}}
\renewcommand{\subsection}[1]
{\vspace*{0.6cm}\addtocounter{subsectionc}{1}
        \setcounter{subsubsectionc}{0}\noindent
       {\normalsize\it\thesectionc.\thesubsectionc.
#1}\par\vspace*{0.4cm}}
\renewcommand{\subsubsection}[1]
{\vspace*{0.6cm}\addtocounter{subsubsectionc}{1}
        \noindent
{\normalsize\rm\thesectionc.\thesubsectionc.\thesubsubsectionc.
        #1}\par\vspace*{0.4cm}}

\newcounter{appendixc}
\newcounter{subappendixc}[appendixc]
\newcounter{subsubappendixc}[subappendixc]

\renewcommand{\appendix}[1] {\vspace*{0.6cm}
        \refstepcounter{appendixc}
        \setcounter{figure}{0}
        \setcounter{table}{0}
        \setcounter{equation}{0}
        \renewcommand{\thefigure}{\Alph{appendixc}.\arabic{figure}}
        \renewcommand{\thetable}{\Alph{appendixc}.\arabic{table}}
        \renewcommand{\theappendixc}{\Alph{appendixc}}
        \renewcommand{\theequation}{\Alph{appendixc}.\arabic{equation}}
        \noindent{\bf Appendix \theappendixc #1}\par\vspace*{0.4cm}}

\renewenvironment{thebibliography}[1]
        {\begin{list}{\arabic{enumi}.}
        {\usecounter{enumi}\setlength{\parsep}{0pt}
\setlength{\leftmargin 1.25cm}{\rightmargin 0pt}
         \setlength{\itemsep}{0pt} \settowidth
        {\labelwidth}{#1.}\sloppy}}{\end{list}}

\newcounter{itemlistc}
\newcounter{romanlistc}
\newcounter{alphlistc}
\newcounter{arabiclistc}

\newcommand{\fcaption}[1]{
        \refstepcounter{figure}
        \setbox\@tempboxa = \hbox{\footnotesize Fig.~\thefigure. #1}
        \ifdim \wd\@tempboxa > 6in
           {\begin{center}
       \parbox{6in}{\footnotesize\baselineskip=12pt Fig.~\thefigure.
#1}
            \end{center}}
        \else
             {\begin{center}
             {\footnotesize Fig.~\thefigure. #1}
              \end{center}}
        \fi}

\newcommand{\tcaption}[1]{
        \refstepcounter{table}
        \setbox\@tempboxa = \hbox{\footnotesize Table~\thetable. #1}
        \ifdim \wd\@tempboxa > 6in
           {\begin{center}
       \parbox{6in}{\footnotesize\baselineskip=12pt Table~\thetable.
#1}
            \end{center}}
        \else
             {\begin{center}
             {\footnotesize Table~\thetable. #1}
              \end{center}}
        \fi}

\def\@citex[#1]#2{\if@filesw\immediate\write\@auxout
        {\string\citation{#2}}\fi
\def\@citea{}\@cite{\@for\@citeb:=#2\do
        {\@citea\def\@citea{,}\@ifundefined
        {b@\@citeb}{{\bf ?}\@warning
        {Citation `\@citeb' on page \thepage \space undefined}}
        {\csname b@\@citeb\endcsname}}}{#1}}

\newif\if@cghi
\def\cite{\@cghitrue\@ifnextchar [{\@tempswatrue
        \@citex}{\@tempswafalse\@citex[]}}
\def\citelow{\@cghifalse\@ifnextchar [{\@tempswatrue
        \@citex}{\@tempswafalse\@citex[]}}
\def\@cite#1#2{{$\null^{#1}$\if@tempswa\typeout
        {IJCGA warning: optional citation argument
        ignored: `#2'} \fi}}

 1
 1
 1

\font\ninerm=cmr9

\def\beq{\begin{equation}}
\def\eeq{\end{equation}}
\def\bea{\begin{eqnarray}}
\def\eea{\end{eqnarray}}
\def\bq{\begin{quote}}
\def\eq{\end{quote}}

\def\AP{{\it Ann.Phys.} }

\def\IJMP{{\it Int.J.Mod.Phys.} }

\def\MPL{{\it Mod.Phys.Lett.} }

\def\NC{{\it Nuovo Cimento} }
\def\NP{{\it Nucl.Phys.} }
\def\PL{{\it Phys.Lett.} }
\def\PR{{\it Phys.Rev.} }
\def\PRL{{\it Phys.Rev.Lett.} }

\def\PTP{{\it Progr.Theor.Phys.} }

\def\PZEKF{{\it Pis'ma v Yh.Eksp.Teor.Fiz.} }
\def\PTP{{\it Progr.Theor.Phys.} }

\def\ZP{{\it Z.Phys.} }

\def\gappeq{\mathrel{\rlap {\raise.5ex\hbox{$>$}}
{\lower.5ex\hbox{$\sim$}}}}

\def\lappeq{\mathrel{\rlap{\raise.5ex\hbox{$<$}}
{\lower.5ex\hbox{$\sim$}}}}

\begin{document}
\pagestyle{empty}
\begin{flushright}
{CERN-TH/95-316}
\end{flushright}
\vspace*{5mm}
\begin{center}
{\bf SUPERSYMMETRY AND GRAND UNIFICATION}\\
\vspace*{1cm}
{\bf John ELLIS}\\
\vspace{0.3cm}
Theoretical Physics Division, CERN\\
1211 Geneva 23, Switzerland\\
\vspace*{3cm}
{\bf Abstract}\\
\end{center}
\vspace*{5mm}
\noindent

%
%
%
Supersymmetry and Grand Unification are the two most
promising directions for
physics beyond the Standard Model. They receive indirect experimental
support from the
apparent lightness of the Higgs boson, the values of the gauge couplings
measured at LEP
and elsewhere, and the persistent solar neutrino deficit.
Phenomenological constraints and
theoretical models constrain predictions in interesting ways. All these
ideas may be
embedded in string theory, which is shown by newly-discovered dualities
to possess
previously-unsuspected richness and simplicity.
\vspace*{4.5cm}
\begin{center}
{\it Invited Rapporteur Talk at the Internal Symposium on}\\
{\it Lepton and Photon Interactions at High Energies}\\
{\it Beijing, August 1995}
\end{center}
\vspace*{1.5cm}
\begin{flushleft}
CERN-TH/95-316\\
November 1995
\end{flushleft}
\vfill\eject
\setcounter{page}{1}
\pagestyle{plain}
\noindent{\bf Beyond the Standard Model}

Although the Standard Model (SM) is in perfect agreement with (almost)
all
experimental data, theorists are not content with it and believe that
something must
lie beyond it. It is common to categorize the open problems left by the
SM into the
{\bf problem of unification}, which motivates the search for a simple
gauge theory that
contains all the gauge forces, {\bf the problem of flavour}, namely why
are there so many
different types of quarks and leptons, and what explains their weak
mixing and CP
violation, and the {\bf problem of mass}. This includes not only the
question of the
origin of the particle masses, to which the SM answer is an elementary
Higgs boson, but
also why all the SM particle masses are so small, to which one possible
answer may be
provided by supersymmetry, as we shall discuss in the rest of this talk.
All these
problems should be resolved in a Theory Of Everything (TOE) which
includes gravity and
reconciles it with quantum mechanics. Such a theory should also explain
the origin of
spacetime, why we live in four dimensions and many other fundamental
problems of
particle physics and cosmology. The only candidate we have for such a
TOE is the
superstring, which will also be discussed at the end of this talk.

\section{Motivations for Supersymmetry}

Supersymmetry\cite{susy} is a  beautiful theory,
but the motivations for it to appear at accessible energies are related
to the problem
of mass mentioned above, namely the origin of the hierarchy of mass
scales in physics,
and its naturalness in the presence of radiative corrections\cite{hier}.
The question
why $m_W$ is much less than $m_{\rm Planck}$ or $m_{\rm GUT}$ can be
rephrased as a
question: Why is $G_F \gg G_N$, or even why the Coulomb potential inside
an atom is
much stronger than the Newtonian potential:
\beq
{e^2 \over r} \lappeq G_N \times {m^2 \over r}
\label{a1}
\eeq
This hierarchy is valuable to radiative corrections. We say that a
theory is natural if
the radiative corrections are not much larger than the physical values
of observable
quantities. For example, the leading one-loop correction to a fermion
mass takes the
form
\beq
\delta m_f = 0\left({\alpha\over\pi}\right)~m_f ~\ln \left({\Lambda\over
m_f}\right)
\label{a2}
\eeq
which is not much larger than $m_f$ for any reasonable cut-off $\Lambda
\lappeq  m_P$.

Naturalness is, however, a problem for an elementary Higgs boson, which
in the
electroweak sector of the SM must have a mass
\beq
m_H = 0\left(\sqrt{{\alpha\over\pi}}\right)^{0\pm 1} \times m_W
\label{a3}
\eeq
The one-loop diagrams shown in Fig. 1 lead to ``large" radiative
corrections of the
form
\beq
\delta m^2_H \simeq g^2_{f,W,H} \int^\Lambda {d^4k\over (2\pi
)^4}~{1\over k^2} =
0\left({\alpha\over\pi}\right) \Lambda^2
\label{a4}
\eeq
These are much larger than the physical value $m_H^2$ if the cut-off
$\Lambda$,
representing the scale at which new physics appears, is of order $m_P$
or $m_{\rm GUT}$.

\begin{figure}
\vspace*{13pt}
\vspace*{1.4truein}             
0.2TRUEIN
\fcaption{Quadratically-divergent one-loop diagrams contributing to
$m^2_H$, $m^2_W$.}
\label{fig:one}
\end{figure}

Supersymmetry solves the naturalness problem of an elementary Higgs
boson\cite{hier} by
virtue of the fact that it has no quadratic divergences  and fewer
logarithmic divergences\cite{noren} than non-supersymmetric theories.
The fermion
and boson diagrams shown in Fig. 1 have opposite signs, so that their
net result is
\beq
\delta m^2_{W,H} \simeq - \left({g^2_F\over 4\pi^2}\right)~~(\Lambda^2 +
m^2_F) +
\left( {g^2_B\over 4\pi^2}\right) ~~(\Lambda^2 + m^2_B)~.
\label{1}
\eeq
The leading divergences cancel if there are the same numbers of bosons
and
fermions, and if they have the same couplings $g_F = g_B$, as in a
supersymmetric theory. The residual contribution is small if
supersymmetry is
approximately valid, i.e., if $m_B \simeq m_F$:
\beq
\delta m^2_{W,H} \simeq 0 \left({\alpha\over\pi}\right)~~\left\vert
m^2_B -
m^2_F\right\vert
\label{2}
\eeq
which is no larger than $m_{W,H}^2$ if
\beq
\left\vert m^2_B - m^2_F\right\vert \lappeq 1~{\rm TeV}^2
\label{3}
\eeq
This property provides the first motivation for supersymmetry at low
energies.
However, it must be emphasized that this is a qualitative argument which
should be regarded as a matter of taste. After all, mathematically an
unnatural theory is
still renormalizable, even if it requires fine tuning of parameters to
obtain the
correct physical values. A second supersymmetric miracle is the absence
of many
logarithmic divergences: for many Yukawa couplings and quartic terms in
the effective
potential\cite{noren},
\beq
\delta\lambda \propto \lambda
\label{4}
\eeq
which vanishes if the rare coupling $\lambda = 0$.
This means that couplings between light and heavy Higgses, which could
devastate the
hierarchy\cite{gildener}, will not appear via quantum corrections if
they are absent at
the tree level.
The combination of Eqs.
(\ref{1}) and (\ref{4}) means that if $m_W \leq m_P$ at the tree level,
it stays
small in all orders of perturbation theory, solving the naturalness
problem and
providing a context for attacking the hierarchy problem.

obtained
to weigh
soft

The minimal supersymmetric extension of the Standard Model
(MSSM)\cite{MSSM} is
characterized by gauge interactions which are the same as those in the
Standard Model (SM), and Yukawa interactions obtained from a cubic
superpotential which is an analytic function of the left-handed fields
\beq
W = \sum_{L,E^c} \lambda_L~ LE^c H_1 + \sum_{Q,U^c}\lambda_U~ QU^c H_2
 + \sum_{Q,D^c}\lambda_D~QD^c H_1 + \mu H_1H_2
\label{6}
\eeq
The first three terms give masses to the charged leptons, charge-2/3
quarks
and charge-1/3 quarks respectively. Two Higgs doublets are needed in
order to
preserve the analyticity of $ W $ and to cancel triangle anomalies.
This implies the introduction of the fourth term in Eq. (\ref{6}), which
couples
the Higgs supermultiplets. The quartic part of the effective scalar
potential is
determined by the gauge and Yukawa interactions, which leads to
the relations between the physical Higgs boson masses to be discussed
later.

In addition to the above supersymmetric parts of the effective action,
supersymmetry breaking is necessary to obtain $m^2_F \not= m^2_B$, which
is
usually parametrized by soft mass parameters for scalars $m_{0_i}$ and
gauginos
$M_{1/2_a}$, as well as soft trilinear and bilinear coefficients
$A_{ijk}$ and
$B_{ij}$. In much the same way as gauge couplings in conventional GUTs,
these
are subject to renormalization:
\beq
M_{1/2_a} \propto \alpha_a~,\quad\quad \tilde{M}^2_{0_i} - M^2_{0_i} +
C_{ia}
M^2_{1/2_a} + D_i M^2_Z
\label{7}
\eeq
where the coefficients $C_{i_a}$ and $D_{i}$ are
calculable\cite{renorm}. It is often
assumed that the soft supersymmetry breaking parameters are universal at
some
high renormalization scale $Q = M_{GUT}$ or $M_P$:
\beq
M_{1/2_a}\vert_Q = M_{1/2}~,\quad M^2_{0_i} = M^2_0
\label{8}
\eeq
This assumption protects the low-energy theory against flavour-changing
neutral
currents \break
(FCNC)\cite{fcnc}, but it is not necessarily true. For example, there
could
be non-trivial renormalization at scales $M_{GUT} \lappeq Q \lappeq
M_P$, so that:
\beq
(M^2_0)_{\mbox{\boldmath$\overline5$}} \not=
(M^2_0)_{\mbox{\boldmath$10$}}
\label{9}
\eeq
in a context of an $SU(5)$ GUT\cite{gutren}, and/or differences may
emerge when the
GUT degrees of freedom are integrated out, and/or the input parameters
may not be
universal at $Q = M_P$ \cite{moddep}:
\beq
M^2_{O_i} = f_i~({\rm moduli})
\label{10}
\eeq
where ``moduli" is a fancy term for vacuum expectation values in a
string
theory. Some such violations of universality may be consistent with the
FCNC
constraints\cite{fcncok}, particularly for heavier generations.

If one nevertheless assumes universality, different experimental
constraints  can be
combined  to compile the physics reach, both present and
future, as in the $(m_0, M_{1/2})$ plane shown in Fig. 2,  or the $(\mu
, M_{1/2})$ plane
shown in Fig. 3 \cite{lep2}. In each case, the diagonally-shaded regions
are those
excluded by present experimental constraints. Also shown in Fig. 2 are
regions excluded
by theoretical considerations. Both figures show the mass contours for
sparticles that
could be studied with future accelerators such as LEP2 or the LHC.

In addition to the search for supersymmetric particles, a promising
avenue for
probing supersymmetry is the search for supersymmetric Higgs bosons. The
two
complex Higgs doublets required in the MSSM contain eight real degrees
of
freedom, of which three are eaten by the $W^\pm$ and the $Z^0$ to give
them
their masses, leaving five physical Higgs bosons to be discovered.
Three of these $(h, H, A)$ are neutral and two $(H^\pm )$ are charged.
At the
tree level, all their masses and couplings are specified in terms of two
parameters, which may be taken as $m_A$ or $m_h$ and $\tan \beta \equiv
v_2/v_1$. These restrictions follow from the supersymmetric form of the
Higgs
potential, and would imply that $m_h < m_Z$ at the tree level but
 there are important radiative corrections\cite{corrhigg}

\begin{figure}
\vspace*{14cm}             
\fcaption{
Present experimental (shaded) and theoretical (bricked)
constraints in the $(m_0,m_{1/2}$) plane, assuming universal
supersymmetry
breaking\cite{lep2}.}
\end{figure}

\noindent
which depend strongly on
the  mass of the top quark, which is now known\cite{mt} to be large:
\beq
\delta m^2_h  \simeq {3g^2\over 8\pi^2}
{}~~ {m^4_t\over m^2_W}~~\ln \left( {m^2_{\tilde q}\over m^2_t}\right)
\label{11}
\eeq
These raise the upper bound on $m_h$ to as large as 130 GeV, as seen in
Fig. 4
\cite{higgmass}.

\begin{figure}
\vspace*{11cm}             
\fcaption{Present experimental constraints and future LEP2 physics reach
in the
$(\mu, M_2 \equiv m_{1/2}\alpha_2/\alpha_{\rm GUT}$) plane\cite{lep2}.}
\label{fig:three}
\end{figure}

\begin{figure}
\vspace*{13pt}
\vspace*{1.4truein}             
0.2TRUEIN
\fcaption{Upper limit on $m_h$ in the MSSM as a function of $\tan\beta$
for zero
(dashed) and maximal (solid) mixing, assuming $m_{\tilde q} = $1 TeV
\cite{lep2}.}
\label{fig:four}
\end{figure}

Before the inclusion of these radiative corrections, experimentalists at
LEP2
could have been quite sure of finding the lightest neutral
supersymmetric Higgs $h$.
Even with these radiative corrections included, they are still able to
explore a
large fraction of the parameter space, as seen in Fig. 5 \cite{lep2}. We
see here the
importance of increasing the centre-of-mass energy of LEP2 as high as
possible.
The search for supersymmetric Higgs bosons at the LHC has also been
studied
intensively during the past year, and Fig. 6 exhibits the domains of
parameter
space that may be explored by the ATLAS and CMS detectors using various
supersymmetric Higgs signatures\cite{lhc}.  We see from Figs. 5 and 6
that LEP2 and the
LHC between them should be able to explore all of the MSSM parameter
space, at
least if the LEP2 energy reaches 192 GeV as is now being proposed.

\begin{figure}
\vspace*{13pt}
\vspace*{1.4truein}             
0.2TRUEIN
\fcaption{ Reach for Higgs bosons in the MSSM at LEP2 with a
centre-of-mass energy of
192 GeV. The dark shaded regions are excluded theoretically\cite{lep2}.}
\label{fig:five}
\end{figure}

\begin{figure}
\vspace*{13pt}
\vspace*{1.4truein}             
0.2TRUEIN
\fcaption{Reach for Higgs bosons in the MSSM at the LHC\cite{lhc}.}
\label{fig:six}
\end{figure}

\section{Possible Experimental Motivations for \hfill\break
Supersymmetry}

The precision electroweak data from LEP and elsewhere provide two (or
three?)
tentative indications favouring a supersymmetric world view. One is that
they
favour a relatively light Higgs boson\cite{efl}$^,$\cite{light}. For
several years, global
fits have consistently given preferred values $m_H \lappeq$ 300 GeV, and
are highly
consistent with the prediction of the MSSM that $m_h \simeq m_Z \pm 40 $
GeV
\cite{corrhigg}$^,$\cite{higgmass}. Figure 7 shows the $\chi^2$ function
for a recent global
fit in the SM, which yields\cite{efl}
\beq
M_H = 76^{+152}_{-50}~{\rm GeV}
\label{12}
\eeq
The $\chi^2$ obtainable in the MSSM is essentially
identical\cite{eflsusy}, whilst
strongly-interacting Higgs models such as those based on technicolour
have much
larger $\chi^2$ and are disfavoured\cite{xtechni}.

\begin{figure}
\vspace*{13pt}
\vspace*{14cm}             
\fcaption{The values of $\chi^2$ as a function of $M_H$ from a global
fit\cite{efl}
to the precision electroweak data.}
\label{fig:seven}
\end{figure}

The second indication favouring supersymmetry is that measurements
of the SM gauge couplings $\alpha_{1,2,3}$ have for some
time\cite{amaldi}$^,$\cite{costa}
favoured supersymmetric GUTs over the minimal non-supersymmetric GUT,
which
predicts\cite{marciano}:
\bea
\sin^2\theta_W(m_Z)\bigg\vert_{\overline{MS}} &=& 0.208 + 0.004 (N_H -1)
+ 0.006
\ln \left({400~{\rm MeV}\over  \Lambda_{\overline{MS}}(N_f = 4)}\right)
\nonumber \\ &=& 0.214 \pm 0.004
\label{13}
\eea
This tendency has been strongly reinforced by the higher-precision data
recently provided by LEP \cite{renton}. Figure 8 gives an overview of
the present
theoretical and experimental situation. The qualitative success of GUTs
in predicting
$\sin^2\theta_W$ is impressive: it is only when we blow the vertical
scale up
by a factor of 10 that we notice a discrepancy with the minimal
non-supersymmetric GUT prediction in Eq. (\ref{13}), and only when we
blow it up
by a further factor of 10 that we begin to wonder whether the LEP data
on
$\sin^2\theta_W$  may fall below the prediction of a minimal
supersymmetric GUT.
However, it should be emphasized that supersymmetric GUTs contain many
parameters, reducing the precision of their
predictions\cite{ekn}$^,$\cite{acpz}: we
shall return to them later.

\begin{figure}
\vspace*{13pt}
\vspace*{1.4truein}             
0.2TRUEIN
\fcaption{Gee-whizz plot showing how well GUT predictions of
$\sin^2\theta_W$ agree
with the experimental data.}
\label{fig:eight}
\end{figure}

Another experimental effect which has excited much interest recently,
including speculations about supersymmetry, is the possible discrepancy
between LEP measurements and the SM predictions for the rates for  $Z^0$
decays into bottom and charm quarks $R_b$ and $R_c$ \cite{renton}. Some
authors have
investigated whether this possible discrepancy could be accommodated
within the
MSSM, if either supersymmetric Higgs bosons or stops and charginos are
light\cite{susyrb},
just above the mass ranges excluded by direct searches. It is possible
to explain $R_b$,
but it is very difficult to explain the central experimental value of
$R_c$, whose
numerical discrepancy with the SM value is even larger, though a smaller
number of standard
deviations\cite{renton}. Many theoretical models share this lack of
success in explaining
simultaneously $R_b$ and $R_c$, and  the latter would be very surprising
if it were to be
confirmed. My present attitude is to wait and see how these experimental
discrepancies
develop\cite{26a}, and not yet to interpret them
as evidence for supersymmetry. With the resolution of the hierarchy
problem, the
indication of a light Higgs boson and the GUT unification of the gauge
couplings, we may
already have enough motivation for supersymmetry!

\section{Grand Unified Theories}

Now is the time to delve deeper in the guts of GUTs, reviewing the
extent to
which they accentuate the hierarchy problem, studying in more detail the
correlation they provide between the values of $\alpha_s (M_Z)$ and
$\sin^2\theta_W$,  and reviewing their predictions for novel phenomena
such as
proton decay and neutrino masses.

The hierarchy problem reviewed in Section 2 can be restated as the
question:
``Why is the electroweak Higgs boson light?" In the context of the
minimal
$SU(5)$ GUT, this question can be reformulated as: ``Why is
$m_{\mbox{\boldmath$2$}}
\leq
m_{\mbox{\boldmath$3$}} $?", where the subscripts denote the doublet and
triplet components
of the five-dimensional Higgs representations. The enormous separation
between these
masses is done by hand in the minimal $SU(5)$ model\cite{dgs}:
\beq
\left. \matrix{m_{\mbox{\boldmath$3$}} \cr m_{\mbox{\boldmath$2$}}}
\right\}
= m_{\mbox{\boldmath$5$}} \left\{\matrix{+2\cr-3}\right\}
\lambda < 0 \vert V_{\mbox{\boldmath$24$}} \vert 0 >  =
\left\{ \matrix{0(M_{\rm GUT}) \cr 0(M_W)\hfill} \right. \quad
\label{14}
\eeq
which requires inelegant and extreme fine-tuning between the bare and
 $<{\mbox{\boldmath$24$}}>$ contributions $m_{\mbox{\boldmath$5$}}$ and
$-2\lambda < 0\vert V_{\mbox{\boldmath$24$}} \vert 0 >$
to the doublet mass $m_{\mbox{\boldmath$2$}}$.
An improvement is possible in principle in missing-partner
models\cite{misspart}, in
which the triplet Higgs components require large Dirac masses from
couplings with other
triplet fields, but there are no such partners for the doublet fields,
which
therefore remain light. This is an elegant idea, but its realization in
conventional GUTs is very complicated, requiring several large Higgs
representations, such as\cite{mpmodels}
\bea
SU(5): &{\mbox{\boldmath$50$}} + \overline{{\mbox{\boldmath$50$}}} +
{\mbox{\boldmath$75$}} + \ldots \phantom{xxxxxxxxxxxxx} \nonumber \\
SO(10): & 3.{\mbox{\boldmath$16$}} + 2.{\mbox{\boldmath$10$}} +
3.{\mbox{\boldmath$45$}}
+ {\mbox{\boldmath$54$}} + \overline{{\mbox{\boldmath$126$}}}
+ {\mbox{\boldmath$126$}}
\label{15}
\eea
The simplest missing-partner mechanism is that \cite{f5gut} in the
flipped $SU(5)\times
U(1)$ GUT\cite{f5gutold}, in which the GUT Higgses occupy {\bf 10} and
${\bf
\overline{10}}$ representations, and the triplet components of the
five-dimensional
electroweak   Higgs representations couple to triplet components of the
GUT Higgses to
require large Dirac masses.
Examples of flipped $SU(5)\times U(1)$ GUTs have been derived in string
theory\cite{stringf5}. However, the other potential solutions to the
hierarchy problem are
problematic in string models: in general, these do not allow bilinear
mass terms of the
type required in Eq. (\ref{14}),  exotic representations like those in
Eq. (\ref{15}) are
not found\cite{nolarge}, and their pattern of couplings may also be
difficult to arrange.

We now explore in
more detail the supersymmetric GUT relation between $\alpha_s (M_Z)$ and
$\sin^2\theta_W$ \cite{33a}. When one looks more carefully at the
gee-whizz plots of the
gauge couplings in the MSSM meeting at a single Grand Unification scale
around
$10^{16}$ GeV, one finds a possible minor discrepancy with the minimal
supersymmetric GUT as already mentioned in the context of Fig. 8. The
supersymmetric GUT prediction for $\sin^2\theta_W$ can be written in the
form\cite{ekn}
\bea
\sin^2\theta_W (M_Z)\bigg\vert_{\overline{MS}} &=& 0.2029 +
{7\alpha_{em}\over
15\alpha_3} + {\alpha_{em}\over 20 \pi}
\bigg [ -3\ln \left({m_t\over M_Z}\right) + {28\over 3} \ln
\left({m_{\tilde
g}\over M_Z}\right) \nonumber \\
&&- {32\over 3} \ln \left({m_{\tilde W}\over M_Z}\right)
- \ln \left({M_A\over M_Z}\right) - 4\ln \left({\mu\over M_Z}\right) +
\ldots \bigg ]
\label{16}
\eea
which involves many supersymmetry-breaking
parameters.  It is convenient to summarize these  in the lumped
parameter\cite{lump}
\beq
T_{\rm SUSY} \equiv \vert \mu \vert ~~ \left({m^2_{\tilde W}\over
m^2_{\tilde
g}}\right)^{14/19}~~
\left({M^2_A\over\mu^2}\right)^{3/38}~~
\left({m^2_{\tilde W}\over\mu^2}\right)^{2/19}~~
\prod^3_{i=1}~~
\left({m^3_{\tilde l_i} m^7_{\tilde q_i}\over m^2_{\tilde e_i}
m^5_{\tilde u_i}
m^3_{\tilde d_i}}\right)^{1/19}
\label{17}
\eeq
If one further assumes universality at the Grand Unification scale, then
approximately
\beq
T_{\rm SUSY} \simeq \mu \left({\alpha_2\over \alpha_3}\right)^{3/2}
\simeq {\mu \over
7}
\label{18}
\eeq
It should be noted that $T_{\rm SUSY} \sim $ 300 GeV corresponds to
squark
masses around 2~TeV. The prediction (\ref{16}) is to be compared with
the
experimental value\cite{expvalue}
\bea
\sin^2\theta_W (M_Z)_{\overline{MS}} &=&
0.2317 \pm 0.0003  +(5.4\times 10^{-6})~~~(m_H - 100~{\rm GeV}) + \ldots
\nonumber \\
&&-(3.03\times 10^{-5})~(m_t - 165~{\rm GeV}) + \ldots \nonumber \\
\nonumber \\
&=& 0.2312 \pm 0.0003 \quad{\rm for}\quad m_H = 100~{\rm GeV}, ~~ m_t =
180~{\rm GeV}
\label{19}
\eea
where the effects of $M_H$ and $m_t$ have been indicated explicitly, but
there
are additional supersymmetric corrections\cite{36a} which may reach the
per cent level.
This comparison yields\cite{highas}
\bea
\alpha_s (M_Z) > 0.126 & {\rm for} & T_{\rm SUSY} < M_Z \nonumber \\
{\rm or} ~~~~~~~~ > 0.121 & {\rm for} & T_{\rm SUSY} < 300 ~{\rm GeV}
\label{20}
\eea
as seen in Fig. 9, with an error of about 0.0015.

\begin{figure}
\vspace*{13pt}
\vspace*{1.4truein}             
0.2TRUEIN
\fcaption{Minimal supersymmetric $SU(5)$ GUT predictions for
$\alpha_s(M_Z)$
\cite{highas}.}
\label{fig:nine}
\end{figure}

Before concluding that supersymmetric GUTs favour values of $\alpha_s
(M_Z)$
above the present world average, one should recall that there are
important
uncertainties in this minimal supersymmetric GUT analysis. For one
thing, there
are in general important GUT threshold effects, which have been
evaluated as
\beq
\delta_{\rm heavy} = {3\over 10\pi} ~~\alpha_{\rm GUT} ~~\ln
\left({M_{H_3}\over
M_{\rm GUT}}\right) + ({\rm positive~terms})
\label{21}
\eeq
in minimal supersymmetric $SU(5)$ \cite{ekn}$^,$\cite{barhall}, while
$\delta_{\rm heavy}$
may be negative:  \beq
\delta_{\rm heavy} \simeq -4 \%
\label{22}
\eeq
in the $SU(5)$ missing doublet model of Eq. (\ref{15}) \cite{negdelt}
and in flipped
$SU(5)\times U(1)$ \cite{f5delt}. As shown in Fig. 10, the
missing-doublet model is in
better agreement with the data on $\sin^2\theta_W$ and $\alpha_s (M_Z)$
than is the
minimal $SU(5)$ model\cite{betal}. Moreover, there could easily be
modifications of the
unification conditions $\alpha_3 = \alpha_2 = \alpha_1$ due to
non-renormalizable interactions scaled by inverse powers of $m_P$
\cite{41a}, which
might yield an uncertainty
\beq
\Delta \alpha_s (M_Z) = \pm 0.006
\label{23}
\eeq

\begin{figure}
\vspace*{13pt}
\vspace*{1.4truein}             
0.2TRUEIN
\fcaption{The missing-doublet model\cite{mpmodels} provides GUT
threshold
corrections $\epsilon_g$ \cite{negdelt} that are in better agreement
with the data
\cite{highas} than is the minimal supersymmetric $SU(5)$ GUT\cite{dgs}.}
\label{fig:ten}
\end{figure}

In view of all these uncertainties, I take the point of view that
supersymmetric GUTs are
still in very good shape, whereas it should be repeated that minimal
non-supersymmetric
GUTs are unquestionably in disagreement with the measured values of
$\sin^2\theta_W$
and $\alpha_s (M_Z)$.

\section{Baryon decay}

As is well known, in  minimal non-supersymmetric $SU(5)$ the preferred
nucleon decay modes are:
\bea
&&p\rightarrow e^+\pi^0~,\quad e^+\omega~, \quad \bar\nu \pi^+~,\quad
\mu^+ K^0~,\ldots
\nonumber \\
&&n\rightarrow e^+\pi~,\quad e^+\rho^-~,\quad \bar\nu \pi^0~,\ldots
\label{24}
\eea
and the best available numerical estimate of the proton lifetime
is\cite{latt}
\beq
\tau (p\rightarrow e^+\pi^0)\simeq (1.4\pm 0.3)\times 10^{32\pm 1}
\times\left(
{M_{\rm GUT}\over 6\times 10^{14}~{\rm MeV}}\right)^4
\label{25}
\eeq
which is to be compared with the present experimental limit\cite{pdg}
\beq
\tau (p\rightarrow e^+\pi^0) > 5.5 \times 10^{32} y
\label{26}
\eeq
In view of the trend for higher-energy measurements\cite{lepas} to find
larger values of
$\Lambda_{\overline{MS}}^{N_f=4}$, which could be as large as 400 MeV
corresponding to
$m_{\rm GUT} \simeq (4$ to 8) $\times 10^{14}$ GeV, I no longer consider
the
conflict between Eqs. (\ref{25}) and (\ref{26}) to be conclusive.
However,
minimal non-supersymmetric GUTs are nevertheless excluded by the
$\sin^2\theta_W$
argument discussed above.

The Grand Unification scale $m_{\rm GUT}$ is increased to
about $10^{16}$ GeV in minimal supersymmetric $SU(5)$, yielding a
lifetime
for proton decay into $e^+\pi^0$ far beyond the present experimental
limit.
However, dimension-five operators in this model yield the alternative
decays
$p\rightarrow \bar\nu K^+, \quad n \rightarrow \bar\nu K^0$ \cite{dim5},
for which the
present experimental limits are less stringent than Eq. (\ref{26})
\cite{pdg}
\beq
\tau (p,n \rightarrow \bar\nu K) \gappeq 10^{32} y
\label{27}
\eeq
The limit (\ref{27}) is (barely) compatible with theory for $M_X
\lappeq
10^{16}$ GeV \cite{susypdk}. Missing-partner
models\cite{misspart}$^,$\cite{mpmodels}
including flipped $SU(5)\times U(1)$ \cite{f5gut} greatly suppress
dimension-five
operators, which are no longer a problem. In the specific case of
flipped $SU(5)\times
U(1)$, the grand unification scale may be somewhat below 10$^{16}$ GeV
\cite{f5delt},
particularly if one takes the lower end of the presently-allowed range
of $\alpha_s(M_Z)$,
in which case $p\rightarrow e^+\pi^0$  and related decays may occur at
observable rates,
though with branching ratios different from minimal non-supersymmetric
$SU(5)$
\cite{aspects}.  Therefore, the Superkamiokande detector about to start
next year may
finally be able to reassure us that protons are not forever!

\section{Neutrino masses and oscillations}

There is no good reason why neutrino masses should vanish, and grand
unified
theorists certainly expect them to be non-zero. The simplest form of
neutrino
mass matrix is the see-saw\cite{gmby}
\beq
(\nu_L~,\quad \bar\nu_R)~~\left(\matrix{m^M & m^D\cr m^D & M^M}\right)~~
\left(
{\nu_L\atop \bar\nu_R}\right)
\label{28}
\eeq
 where $\nu_R$ is a singlet right-handed neutrino field, and
\beq
m^D = g_{H\bar\nu\nu} < 0\vert H_{\Delta I = 1/2} \vert 0 >
\label{29}
\eeq
 is a generic Dirac mass which is of order  the charge-2/3 quark mass
$m_{2/3}$
in many models, and $m^M,  M^M$ are  $\Delta I = 1, 0$ Majorana masses
which are
expected to be of order $M_W^2/ M_X, M_X$, respectively.
When one diagonalizes the matrix (\ref{28}), one finds mass eigenstates
of the
generic form
\bea
\nu_L + 0\left({m_W\over m_X}\right)\bar\nu_R &:& m = 0\left({M^2_W\over
M_X}\right) \nonumber \\ \nonumber \\
\nu_R + 0\left({m_W\over m_X}\right)\bar\nu_L &:& M = 0\bigg( M_X \bigg)
\label{30}
\eea
where ``$M_X$" should be understood as anywhere between $m_P$ and
$O(\alpha/\pi)^2 m_{\rm GUT}$, depending on the model. Generically,
(\ref{28}),
(\ref{29}) and (\ref{30}) yield the guess that
\beq
m_{\nu_i} \sim {m^2_{{2\over 3}_i} \over M_{X_i}}
\label{31}
\eeq
for the three generations $i = 1, 2, 3$ of light neutrinos.

There are of course many more complicated models of neutrino masses
incorporating more fields and/or more couplings, but this simple see-saw
model
accommodates in a very natural way the apparent deficit of solar
neutrinos\cite{solnu},
and correlates it with the astrophysical wish for a hot Dark Matter
particle\cite{mdm}.
In my view, it is becoming increasingly difficult to retain an
astrophysical
explanation for the solar neutrino deficit, particularly in view of the
strengthening helioseismological constraints  on the solar model,
including its
central temperature. As reviewed here by Winter\cite{winter}, the most
appealing
interpretation of  the solar neutrino deficit invokes matter-enhanced
neutrino
oscillations\cite{msw}:
\beq
\nu_e\rightarrow \nu_{\mu~{\rm or}~ \tau} ~:~\Delta m^2 \sim
10^{-5}~{\rm ev}^2~,
\quad \sin^2 2\theta \sim \left\{ \matrix{~~~10^{-2} \cr  {\rm or}\hfill
\cr
{}~~~1\hfill}\right. \label{32}
\eeq
Theoretical prejudice (\ref{31}) and the small values of
inter-generational
mixing angles observed in the quark sector favour the scenario
\beq
m_{\nu_e} \ll m_{\nu_\mu} \sim  3\times 10^{-3}~{\rm eV} \ll
m_{\nu_\tau} ~, \quad
\sin^2 2\theta_{e\mu} \sim  10^{-2}
\label{33}
\eeq
Scaling the inferred value of $m_{\nu_\mu}$ by  $m_t^2 / m_c^2$ and
allowing $M_2/M_3
\sim 1/10$ leads naturally to the guess that
\beq
m_{\nu_\tau} \sim 7~{\rm eV}
\label{34}
\eeq
as favoured in mixed dark matter models of cosmological structure
formation,
and
\beq
\sin^2 2\theta_{\mu\nu} \sim 10^{-0(3)}
\label{35}
\eeq
which may be accessible to the new generation of accelerator neutrino
oscillation experiments, CHORUS and NOMAD at CERN, and E803 at
Fermilab\cite{mdm}.

As reviewed here by Winter\cite{winter}, there are other suggestions of
mass and
oscillation effects in atmospheric neutrinos\cite{kamatmos} and the LSND
experiment\cite{lsnd}, but I prefer to wait and see whether these claims
become confirmed.

\section{Further Dynamical Ideas}
\subsection{Electroweak Symmetry Breaking}

It has been suggested\cite{ir} that the breaking of electroweak symmetry
may be driven
by renormalization of the soft supersymmetry breaking
parameters\cite{renorm} introduced
earlier. This renormalization may resolve the apparent
conflict between the preference of the super-Higgs mechanism  for
generating
$m_0^2 > 0$ with the requirement that $m_H^2 < 0$ for the electroweak
Higgs
mechanism. The dominant renormalization effects are those due to gauge
couplings and the top Yukawa coupling, which have opposite signs.
If one follows the renormalization down to sufficiently low scales $Q$,
large top Yukawa coupling may drive $m_H^2 (Q) < 0$, triggering $m_W
\not= 0$ \cite{ir}.
This occurs at a scale $Q$ hierarchically smaller than the input scale,
so that
\bea
{m_W\over m_P} = \exp \left( {-0(1)\over \alpha_t}\right) &;& \alpha_t =
{\lambda^2_t\over 4\pi} \nonumber \\ \nonumber \\
&& m_t = \lambda_t < H >
\label{37}
\eea
Typical dynamical calculations\cite{ir} yield $m_t$ in the range now
found by
experiment.

\subsection{Supersymmetry Breaking}
The above mechanism for electroweak symmetry breaking requires soft
supersymmetry breaking to be put in $a~priori$: It is also possible that
the
scale of supersymmetry breaking may be determined by quantum
effects\cite{elnt}. Consider,
for example, a model with no potential at the tree level in some flat
direction
in the space of moduli\cite{cfkn}, so that it is independent of the
generic supersymmetry
breaking scale $\tilde m$:
\beq
{\partial V_{eff}\over \partial \tilde m} = 0
\label{38}
\eeq
One then calculates the quantum corrections to the potential, which
include the
following terms at the one-loop level:
\beq
\delta V_{eff} \ni (\sum_B - \sum_F)\Lambda^4, (\sum_B -
\sum_F)M^2\Lambda^2,
 (\sum_B - \sum_F) m^4 \ln {m^2\over\Lambda^2}
\label{39}
\eeq
where $\Lambda$ is a cut-off scale which we may identify with $M_P$. The
first
term is absent in any supersymmetric theory, since the numbers of bosons
and
fermions are equal. The second term may be absent in specific
supergravity or
superstring models\cite{ekon}$^,$\cite{strm2}. Assuming that this is the
case\cite{62a}, the
effective potential  enables
the supersymmetry breaking scale and hence $M_W$ to be determined
dynamically\cite{62b}.

Another suggestion is that supersymmetry breaking may occur
non-perturbatively
in a hidden sector of the theory, triggered by gaugino
condensation\cite{gaugino}. It is
even possible to imagine mechanisms which combine features of both of
these
scenarios.
There are also ideas that, even within a fixed overall scale of
supersymmetry
breaking, the ratios of supersymmetry breaking parameters, i.e., the
internal
direction in super-flavour space, may be determined dynamically by
radiative
corrections\cite{plastic}.

\subsection{Quark and Lepton Masses}
The next step in a programme of determining dynamically all light mass
scales is
to tackle the fermion mass problem. For example, in many  superstring
models, the top mass is given by
\beq
m_t = \lambda_t < H_2 > ~:~ \lambda_t = g \times f~ ({\rm moduli})
\label{40}
\eeq
where $g$ is the gauge coupling and the moduli (vacuum parameters) may
include
radii of compactification and other quantities which are to be treated
as
quantum fields. These moduli are also often undetermined at the tree
level. Perhaps
these are also determined by quantum corrections, in much the same way
as $m_W$ and
$\tilde m$ \cite{dynmass1}. Such a scenario can be developed not only
for determining
$m_t$, but also $m_b$ and $m_\tau$ \cite{dynmass2}.

\section{The Constrained MSSM}

It is apparent from the preceding discussion that the MSSM contains many
parameters beyond those already present in the SM: $m_{0_i}, M_{1/2_a},
\mu ,
\tan\beta, A_{ijk}, B_{ij}, \ldots$. In an attempt to reduce the
dimensionality of this
parameter space, it is desirable to impose necessary (plausible)
phenomenological and theoretical constraints, which may include the
following:
\begin{itemize}
\item  No sparticles seen: We know from LEP1 that\cite{pdg}
\beq
m_{\tilde l}~,\quad m_{\chi^\pm} \gappeq 45~{\rm GeV}
\label{41}
\eeq
 and from the Fermilab $p\bar p$ collider that\cite{fnalex}
\beq
m_{\tilde q}~,\quad m_{\tilde g} \gappeq 150~{\rm GeV}
\label{42}
\eeq
\item  No Higgs bosons seen: We know from LEP that\cite{pdg}
\beq
m_{h,A} \gappeq 50~{\rm GeV}
\label{43}
\eeq
\item   Small FCNC: As mentioned earlier, this occurs naturally if the
$m_{0_i}$
are universal \cite{fcnc}, but this assumption is not
necessary\cite{fcncok}.
\item  $b\rightarrow s\gamma$: The fact that this decay  has been seen
at a
rate close to that predicted in the SM constrains MSSM
parameters\cite{bsg}. If there are
no light sparticles, this constraint places a stringent lower bound on
$m_{H^\pm}$, which may, however, be relaxed if some other sparticles are
light.
\item $\mu\rightarrow e\gamma$: This is not such a stringent constraint
at the
present time, but might become so in the future\cite{70a}.
\item $g_{\mu} -2$: The forthcoming BNL experiment\cite{bnlgmu} should
impose significant
constraints on the sparticle spectrum\cite{gmu} when it achieves its
designed
sensitivity.
\item Neutron electric dipole moment: This imposes important constraints
on
possible CP-violating phase parameters in the MSSM\cite{edm}, which
depend on the overall
sparticle mass scale.
\item Cold Dark Matter density: The lightest supersymmetric particle is
a good
candidate for Cold Dark Matter, since $R$-parity guarantees its
stability in
many models, and its relic density lies in the desired range
\beq
0.1 \gappeq \Omega_\chi h^2 \gappeq 1
\label{44}
\eeq
for generic values of the parameters\cite{ehnos}.
The resulting constraints on the MSSM are quite sensitive to the
magnitude of
CP violation\cite{cpcdm}.
\end{itemize}

One may add to the above phenomenological constraints some theoretical
constraints, which are more speculative and hence more interesting.
These
include dynamical electroweak symmetry breaking\cite{ir}, possibly
supplemented by some
no-fine-tuning requirement\cite{nft}:
\beq
{\Delta M_W \over M_W} \lappeq \eta_i~~{\Delta~I_i\over I_i}
\label{45}
\eeq
where $I_i$ is some generic input parameter, and $\eta_i$ parametrizes
the amount of
fine tuning. The absence of fine tuning was the basic phenomenological
motivation for
supersymmetry introduced in Section 1, but it is a matter of taste how
to quantify it:
should $\eta_i$ be less than 1?  10? 100? 10$^5$? This argument
certainly favours $m_0,
M_{1/2} \lappeq$ a few hundred GeV \cite{nftmass}, as seen in Fig. 11.

\begin{figure}
\vspace*{13pt}
\vspace*{1.4truein}             
0.2TRUEIN
\fcaption{ Fine-tuning upper limits on the possible sparticle spectrum
assuming
universal (dashed, solid) or non-universal (dash-dotted) squark
masses\cite{fcncok}.}
\label{fig:eleven}
\end{figure}

 One might also postulate the dynamical determination
of other scales, such as $\tilde m, m_t, \ldots $ as discussed above, or
constraints
arising from an infrared fixed-point analysis\cite{stringx}. One may
also impose some
string-motivated Ansatz for supersymmetry breaking, such as
\beq m_{1/2_a} = A = B =
m_{3/2}~,\quad m_{0_i} = 0
\label{46}
\eeq
at the string input scale\cite{stringx}. This type of game is very
exciting and
predictive, but one should always remember that
\beq
{\rm Prob~(Result)} = \prod^{\infty}_{i=1}~{\rm Prob~(Assumption)}_i
\label{47}
\eeq
After expressing these words of caution, let us now look at some
examples of
constrained MSSM calculations.

Figure 12 shows an example\cite{cmssm1} in which the measured value of
$m_t$ favours two
possible solutions, one with small $\tan\beta$ and one with large
$\tan\beta$. The large $\tan\beta$ solution has the additional
attractive
feature that it can accommodate equality between the $t$ and $b$ Yukawa
couplings, as favoured in some string models.  The two solutions yield
different preferred ranges of sparticle masses, as seen in Fig. 12.
Another
example of a MSSM scenario\cite{cmssm2} is shown in Fig. 13, where the
possible masses of
the sparticle species are plotted as a function of the lightest chargino
mass. In this scenario, the right-handed sleptons have only barely
escaped
detection at LEP1 and the lightest chargino should be discovered at
LEP2, as
should the lightest supersymmetric Higgs boson.
This scenario also suggests that the $p\bar p$ collider at Fermilab may
be
able to see  dilepton and trilepton events
due to sparticle pair production and decay.

\begin{figure}
\vspace*{13pt}
\vspace*{1.4truein}             
0.2TRUEIN
\fcaption{ Results from a constrained MSSM\cite{cmssm1}, indicating two
preferred
regions at small and large $\tan\beta$, the latter being consistent with
equal $t$- and
$b$-quark Yukawa couplings.}
\label{fig:twelve}
\end{figure}

\begin{figure}
\vspace*{13pt}
\vspace*{1.4truein}             
0.2TRUEIN
\fcaption{ Results for the sparticle spectrum in a constrained
MSSM\cite{cmssm2}. }
\label{fig:thirteen}
\end{figure}

\section{String Theory}

This is the only candidate we have for a Theory of Everything (TOE). It
is an
apparently consistent quantum theory of gravity, at least at the
perturbative
level and possibly also non-perturbatively. It provides a framework for
tackling the thorny issues of space-time foam, cosmology, the
cosmological
constant, etc.  It also provides a framework for unifying the particle
interactions. However, whereas initially it was thought that there might
be a
unique string model, namely the $D = 10$ $E_8$ heterotic
string\cite{het}, or perhaps
only a few models, subsequently many consistent string models have been
found.
These include a multitude of apparently consistent compactifications of
the
original heterotic string\cite{comphet}, but the most general
formulation of such models
is as heterotic strings directly in four dimensions\cite{d4het}. These
different models
may be regarded as different vacua, i.e., solutions of the classical
equations for the
moduli, of the same underlying string theory. All couplings correspond
to expectation
values of fields (moduli), for example for the gauge couplings $g_i$:
\beq
g^2_i = {k^2_i\over <S>}
\label{48}
\eeq
where the $k_i$ are  Kac-Moody level parameters to which we return
later, and
$S$ is a type of dilaton field.

In all this
confusing thicket of string models, one can make some generic
predictions. For
example, the string unification scale at which $\alpha_i = \alpha_j =
\alpha_{\rm graviton}$ can be predicted\cite{msu}
\beq
m_{SU} \simeq 5 \times 10^{17} g_{\rm GUT}~~{\rm GeV}
\label{49}
\eeq
There is also a generic prediction for $m_t$,
as mentioned earlier
\beq
\lambda_t = g_{\rm GUT} \times f~({\rm moduli})
\label{50}
\eeq
which leads to the qualitative expectation that $m_t/M_W = O(1)$, with
the
possibility of dynamical determination discussed earlier.

Among the techniques used in string model building are the
compactifications of
the $D = 10$ heterotic string\cite{comphet} mentioned earlier,
orbifolds\cite{orb}, free
fermions on the world sheet\cite{ffws},
etc., all of which have been used to produce models with gauge groups of
the
form $SU(3)\times SU(2)\times U(1)^n$ \cite{u1models}. Making a string
GUT is more
problematic, because these typically require adjoint Higgs
representations (e.g., the {\bf
24} of $SU(5)$), which are not available if we maintain space-time
supersymmetry
and restrict ourselves to the level $k_i = 1$ \cite{nolarge}. This was a
motivation for
resuscitating flipped $SU(5)\times U(1)$ \cite{f5gut}, which, as
discussed earlier, also
has an elegant missing-partner mechanism, a see-saw neutrino mass
matrix, and proton decay
at a rate which may be accessible to Superkamiokande if $\alpha_s (M_Z)$
is in the lower
half of the range presently allowed by experiment.
However,
this and other string models lose (or at least weaken) the minimal
supersymmetric GUT prediction for $\sin^2\theta_W$. For this and other
reasons,
theorists have been trying to construct supersymmetric $SU(5)$ and
$SO(10)$
GUTs using higher-level Kac-Moody algebras\cite{stringgut}. The models
found so far either
have more than three generations\cite{moregen}
or other additional chiral stuff\cite{chiral}, but developments in this
quest are very
promising and should be watched.

Let us turn finally to a dramatic new development in string theory,
which may
diminish significantly the apparent proliferation of string models. As
discussed here by Rubakov\cite{rubakov}, it has recently been realized
that gauge theories
with extended supersymmetries have\cite{seiwitt}  amazing duality
properties\cite{montol},
which interrelate strong- and weak-coupling descriptions of the same
physics. It has also
been realized that string theories  possess many such duality
properties\cite{dualrev}.
These include so-called $T$ duality, of which the simplest example is
the equivalence
between a string compactified on a loop of radius $R$ and one
compactified on a loop of
radius $1/R$. This symmetry relating different moduli is believed to be
elevated to a
symmetry at least as large as $SL(2,Z)$. String theory may also possess
an $S$ duality
interrelating strong and weak coupling $<S> \leftrightarrow 1/<S>$ in
Eq.
(\ref{48}), which may also be elevated to $SL(2,Z)$ \cite{sdual}. Even
more excitingly,
many examples have been found of string-string dualities, namely
equivalences
between different types of string, one weakly coupled and one strongly
coupled. Figure 14 is a provisional map of some string dualities, which
apparently
include,  for example, an equivalence between the $D = 10$ heterotic
string
compactified on a four-dimensional torus $T_4$ and the type IIA string
compactified on a $K_3$ manifold\cite{example1}, as well as many others.
One of the most
striking dualities is that between the heterotic $SO(32)$ string and the
type
I $SO(32)$ string\cite{example2}, with spinors of the former interpreted
as solitons of
the latter, and the type IIB string appears to be
self-dual\cite{selfdual}. There are also
duality symmetries\cite{example2} relating string theories with $D = 11$
supergravity\cite{dllsngra} and with supermembrane theories!
\cite{membrane}

\begin{figure}
\vspace*{13pt}
\vspace*{1.4truein}             
0.2TRUEIN
\fcaption{ A provisional map of some of the string dualities recently
discovered.}
\label{fig:fourteen}
\end{figure}

This is a
rapidly-moving field with many new results being obtained\cite{dualrev}.
It offers the
possibility that many different types of string model may simply be
re-expressions of the same underlying theory, whose most basic
formulation may well lie
beyond the concept of string. Any such development could only comfort
the belief that we
have found the TOE.

\section{Conclusions}

There are good theoretical and experimental motivations to hope that we
are
finally on the brink of discovering new physics beyond the Standard
Model.
Precision data from LEP1 and elsewhere suggest that the Higgs boson is
light, in agreement with the prediction of supersymmetry, and may well
be
accessible to LEP2. The consistency between measurements of $\alpha_s
(M_Z)$ and $\sin^2\theta_W$ and the predictions of supersymmetric GUTs
is
certainly encouraging, even if it does not yet enable us to determine
with
any accuracy the scale of supersymmetry breaking. The persistent solar
neutrino deficit seems ever more difficult to explain using
astrophysics,
and may be the harbinger of neutrino masses and oscillations.

The exploration of large new domains of supersymmetry and GUT parameter
space
is about to start, with the advent of LEP2, a new generation of
accelerator
neutrino oscillation experiments pioneered by CHORUS and NOMAD, and a
new
generation of large underground experiments pioneered by Superkamiokande
and
SNO. Will our luck finally change? Will the next
meeting in this series become the first Slepton-Photino Symposium?

\end{document}